%% file: paper.tex
\documentclass[10pt,conference,compsocconf]{IEEEtrans}

 \usepackage[colorlinks,linkcolor=blue,citecolor=blue,urlcolor=blue]{hyperref}
\usepackage{tabularx}
\usepackage{url}
\usepackage{balance}
\usepackage[normalem]{ulem}

\usepackage{eso-pic}
\makeatletter
   \AddToShipoutPicture{%
     \put(110,20){To appear in the Proceedings of the International Conference on Software Engineering, 2012.}
 }
\makeatother

\begin{document}

\title{Semi-automatically Extracting FAQs to Improve Accessibility of \\Software Development Knowledge}

\author{\IEEEauthorblockN{Stefan Hen\ss}
\IEEEauthorblockA{Technische Universit\"at Darmstadt\\
Darmstadt, Germany\\
stefan.henss@gmail.com}
\and
\IEEEauthorblockN{Martin Monperrus}
\IEEEauthorblockA{
University of Lille \& INRIA\\
Lille, France\\
martin.monperrus@univ-lille1.fr}
\and
\IEEEauthorblockN{Mira Mezini}
\IEEEauthorblockA{Technische Universit\"at Darmstadt\\
Darmstadt, Germany\\
mezini@st.informatik.tu-darmstadt.de}
}

\maketitle

\input{abstract}

\section{Introduction}
\input{introduction}

\section{Overview}
\label{overview}
\input{overview}

\section{Data and Pre-Processing}
\label{data-preprocessing}
\input{data-preprocessing}

\section{Topic Mining}
\label{topic-mining}
\input{topic-mining}

\section{FAQ Assembly}
\label{faq-assembly}
\input{faq-assembly}

\section{Qualitative Evaluation}
\label{qualitative-evaluation}
\input{qualitative-evaluation}

\section{Quantitative Evaluation}
\label{quantitative-evaluation}
\input{quantitative-evaluation}

\section{Related Work}
\label{related-work}
\input{related-work}

\section{Conclusion}
\label{conclusion}
\input{conclusion}

\section*{Acknowledgement}
This work was supported by the German Federal Ministry of Education and Research (BMBF) within EC SPRIDE.

\bibliographystyle{ieeetr}
\balance\bibliography{paper}

\end{document}

%% file: abstract.tex
\begin{abstract}
Frequently asked questions (FAQs) are a popular way to document software development knowledge. As creating such documents is expensive, this paper presents an approach for automatically extracting FAQs from sources of software development discussion, such as mailing lists and Internet forums, by combining techniques of text mining and natural language processing. We apply the approach to popular mailing lists and carry out a survey among software developers to show that it is able to extract high-quality FAQs that may be further improved by experts. 
\end{abstract}

%% file: introduction.tex
The term \textit{Frequently Asked Questions} (FAQ) refers to a documentation format which lists questions, as they are, or might be, asked by the target audience, and the corresponding expert answers (also known as Q\&As).
In software development, this format is used by many projects as part of their documentation, for example Linux\footnote{\url{http://tldp.org/FAQ/Linux-FAQ/index.html}}, Apache Lucene\footnote{\url{http://wiki.apache.org/lucene-java/LuceneFAQ}} and Eclipse SWT\footnote{\url{http://www.eclipse.org/swt/faq.php}}. 
Even entire books on software development are written in the FAQ format \cite{Arthorne2004}.

With their typical characteristics, FAQs are complementary to other kinds of software documentation.
Reference manuals are meant to be comprehensive and well-structured but are not suited for providing independent pieces of knowledge targeting practical problems.
Mailing lists and forums do provide such pieces of knowledge but suffer from the sheer mass of information, which makes them difficult to use for novices.
Even despite voting and tagging facilities that can be found in advanced social media sites like stackoverflow.com, good Q\&A pairs are still lost in myriads of messages and can not easily be found \cite{Gottipati2011}.
Furthermore, social media channels often cannot replace first-hand documentation, especially for commercial software (see for instance the FAQ on Microsoft Silverlight\footnote{\url{http://www.microsoft.com/silverlight/faq/}}).
As a result, there is a place for FAQs to fill the gap between traditional documentation and Web 2.0 user support.

As with other kinds of software documentation, creating high quality FAQs is an intellectually challenging and time consuming endeavor, especially since it requires foreseeing potential questions of the reader.
The question we raise in this paper is whether this endeavor can be facilitated by automatically generating parts or even whole FAQs.
This relates to other software engineering research aiming at generating software documentation (e.g. \cite{Long:2009:AHV:1595696.1595727,Bruch2010,Sridhara:2010:TAG:1858996.1859006}), but to our knowledge, the idea of generating FAQs is novel.

Our key insight is that one could extract FAQs from support channels such as mailing lists or Q\&A communities by identifying popular, reoccurring topics (e.g., compiler errors) and finding representative Q\&A pairs for them.
To do so, we mine topic models using latent Dirichlet allocation (LDA) \cite{Blei:2003:LDA:944919.944937} complemented with pre-processing and post-processing steps specifically designed for software development discussions. 
From those topic models, we then assemble the final FAQs by extracting question-answer pairs that closely relate to the mined topics and further fulfill some quality heuristics.
The whole approach is solely based on the discussions and does not require any metadata.

While the FAQ creation itself is automated, the generated FAQs are meant to be read by human users and endorsed by software product managers.
Hence, the generated Q\&As may be validated and edited (e.g., reformulation and clarification) before publishing.
To this extent, our approach to generating FAQs is \emph{semi-}automated.

To evaluate the approach, we have generated FAQ candidates for 50 major open source projects referenced at ohloh.net and invited their lead software developers to assess them.
Among other results, this evaluation shows that there is a clear correlation between our relevance- and quality-heuristics and the experts' approval of our question selections and that 82\% of the answers selected by our system are correct.
Furthermore, we provide quantitative arguments on the system's ability to mine meaningful topic models from software discussions.
The results of those evaluations confirm that FAQs can automatically be distilled from raw knowledge available in software  development discussions.

The rest of this paper is structured as follows:
Section \ref{overview} gives an overview of the approach.
Section \ref{data-preprocessing} explains our data model and how we pre-process the raw input data.
From this data we then mine topic models using LDA as described in Section \ref{topic-mining}.
The resulting models serve as references when assembling the final FAQs using the techniques presented in Section \ref{faq-assembly}. 
The results of our survey involving open source leaders is presented in Section \ref{qualitative-evaluation}, the quantitative evaluation of the approach in Section \ref{quantitative-evaluation}.
Finally, we discuss related work in Section \ref{related-work} and conclude in Section \ref{conclusion}.

%% file: overview.tex
Our approach consists of three phases.
First, conversation entries obtained from mailing lists, forums, Q\&A communities and alike are prepared to be displayed in a FAQ.
Raw conversations, especially from the Internet, generally contain a significant amount of noise, due to how they are presented (e.g. HTML markup), to annotations within the content (e.g. reply information ``[...] wrote: [...]'') and to social content (e.g. ``Best regards, [...]").
As we will show later, this noise also significantly impacts the performance of our algorithm, especially topic mining.
Hence, to obtain meaningful, well-shaped results, we define several pre-processing heuristics to remove noise specific to software development discussions.

Second, for extracting focused FAQs, latent Dirichlet allocation (LDA) is used to identify clusters of related conversations, which, in the following, will be called ``topic models'' or just ``topics''.
LDA characterizes topics through a bag of words (their terminology) and for each conversation, LDA estimates a probability distribution over the mined topics.
In our context, each topic model, depending on its granularity (as determined by how many topics are to be mined), either represents a general category (e.g. configuration) or a specific, reoccurring issue (e.g. a bug).

Finally, in the third phase of the approach, the topic models are used to obtain the final FAQs.
We define means of selecting relevant topics, means for selecting precise questions and good answers from the conversations associated with a topic, as well as means of ordering the final Q\&As.

Note that even if the term "frequent" in "FAQ" suggests the contrary, questions included in a FAQ do not necessarily have to be frequently asked to be valuable for documentation purposes.
If a Q\&A is well referenced by search engines, the question may not be posed many times;  yet, such a Q\&A may be perfectly appropriate for a FAQ.
As a result, the task of creating valuable FAQs is to a lesser extent a matter of measuring question frequencies; it is rather a matter of identifying good Q\&A pairs that generally relate to the concrete problems that users face.

%% file: data-preprocessing.tex
\subsection{Data Model}
In order to generate FAQs from a variety of sources, like mailing lists, Internet forums, newsgroups or Q\&A communities, our approach assumes a very basic data model.
In general, a user starts a new conversation by presenting his issue (the FAQ's question) in a first message and all following replies try to answer it or contribute to a discussion towards the solution.
We assume that the best reply (the FAQ's answer) is a single response of the whole conversation (we do not merge different contents to create an answer).

We pre-process this raw data to create two different ``views'':
the first one aims to be the text appearing in the final FAQs, it must have the ``look'n'feel'' of FAQs; the second one is used for topic mining and data analysis and hence aims at minimizing noise.

\subsection{Look'n'Feel Pre-Processing Heuristics}
\label{rep-display}
The following heuristics filter out most of the content that is irrelevant for FAQ readers.

\emph{Heuristic 1 (Regular expressions):}
First, we use a list\footnote{\url{http://faqcluster.com/regex}} of regular expressions to remove email headers, irrelevant markup (e.g. HTML) and other forms of formatting (e.g. reply quotations such as ``\textgreater What is [...]?'').
Also, certain figures of regular speech, such as greetings and expressions of gratitude, are removed.
In the resulting version, approximately 5\% of all characters are filtered out by this heuristic.

\emph{Heuristic 2 (Determining frequent sentences):}
Second, we empirically observed that sentences (delimited by dot or multiple line breaks) reoccurring at least 10 times within all conversations are unlikely to contain question-specific information (e.g. ``My problem is as follows [...]'').
Those sentences unrelated to a specific issue generally decrease the look'n'feel of resulting Q\&As and also introduce noise to our algorithm.
By applying this heuristic we remove 15\% of sentences within all conversations. 

\subsection{Mining-related Pre-Processing Heuristics}
\label{internal-representation}

 \begin{figure}[!t]
\renewcommand{\arraystretch}{1.2}
\centering
\begin{footnotesize}\begin{tabularx}{\columnwidth}{X|X}
\bfseries FAQ representation & \bfseries Data-mining version\\
\hline
I think the only way to restore the link  is to go into the XML and clean up that clause. & restore link XML clean clause \\
\end{tabularx}
\end{footnotesize}
\caption{The Effect of Heuristic \#3 Illustrated}
\label{fig:technical-representation}
\vspace{-.5cm}
\end{figure} 

In order to maximize the approach's efficiency, we further devised additional heuristics specific to improving the quality of the topics mined by LDA and their post-processing.
However, these heuristics remove information which has to be kept in the final FAQs (as opposed to heuristics \#1 and \#2), e.g., code fragments.
Hence, two linked versions of each conversation entry are stored, one for the data analysis, one for the end-user FAQs, as shown in Figure \ref{fig:technical-representation}. 

\emph{Heuristic 3 (Removing stop words):}
First, we use a list of most frequent English words that are removed from the conversation as, due to their generic use, they carry no relevant information (also known as ``stop words'').
This list also includes  manually added terms which are frequently used in the analyzed domain but known to be not semantically significant for topic mining.
Examples are ``Java'' or ``class'' for discussions about Java software: their high frequency biases and hides the actual topics of the conversations.

Software-related conversations often contain large passages of source code snippets, configuration files, automatically created debug output (e.g. stack traces) and alike.
This can as well lead to topics being heavily distorted by the high concentration of special terminology (e.g., Java keywords) that outnumbers semantically relevant terms.
For example, we found one user submitting large technical outputs including his user name in system paths.
This resulted in one topic being focused on his user name and therefore including most of the conversations containing his name, despite them being related to separated concerns or even different software.

\emph{Heuristic 4 (Removing long paragraphs):}
In our data basis of software-related conversations, very few paragraphs of natural language are longer than 800 characters, while many paragraphs of generated text (e.g., debug output) or source code are much longer.
Filtering those paragraphs, we are able to remove 32\% of all characters.

\emph{Heuristic 5 (Removing paragraphs with too much punctuation):}
Additionally, we found that any paragraph containing more than 200 characters with at least 4\% of them being punctuation (except dots) is very likely to be source code or computer generated (e.g., stack traces, which use to contain many brackets and colons) and we therefore filter them out as well.
10\% of all paragraphs are removed by this heuristic.
Please note that the size limit not only prevents false positives, but also preserves short output like error messages, which rather support topic mining than introducing noise.

The impact of all heuristics is thoroughly evaluated in Section \ref{parameters}.

%% file: topic-mining.tex
We use topic mining to extract topics discussed in software development conversation (e.g., one topic can be related to compilation, while another relates to user-interface customization).
Topic mining enables us (a) to group conversations by common characteristics, and (b) to obtain information about the terms that characterize the topics.
The approach is fully automatic, i.e., the FAQ topics are not required to be sketched beforehand; instead they are extracted from the corpus of conversations.
To mine topics, we chose the latent Dirichlet allocation model which has a mature open-source Java implementation \cite{McCallumMALLET}.
In the following, we use the generic term ``document'' to refer to entire conversations, i.e., all entries of one conversation are merged to a single textual element for LDA to be able to analyze conversations as a whole.

\subsection{Latent Dirichlet Allocation}
\label{LDA}

\begin{figure}
\renewcommand{\arraystretch}{1.2}
\begin{footnotesize}
\begin{tabular}{l}
{\bf Bag of Words for Topic ``Apache Spamassassin''}\\
\hline
spamassassin (.039), spam (.027), mail (.021), rules (.014), dbg (.012),\\ score (.01), email (.01), spf (.009), tests (.009), rule (.008)
\end{tabular}
\end{footnotesize}
\caption{Top 10 words of a topic model representing Apache Spamassassin. The numbers indicate a relative importance (the sum of all words is 1), hence the word ``spamassassin'' is five times more important (expressed through frequency) than ``rule''.}
\label{fig:spamassassin}
\vspace{-.5cm}
\end{figure}

LDA characterizes each topic $i$ as a probability distribution over words (noted $\varphi_{i}$).
For example, conversations related to Java will likely contain terms like ``object" or ``class", hence they will have a high probability in a related model. 
Figure \ref{fig:spamassassin} shows an excerpt of this probability distribution for an actually mined topic related to Spamassassin, a spam filtering software package.
Additionally, since each document can refer to several topics (e.g., related to Java, XML and Eclipse), LDA also defines a probability distribution over topics for each document $j$ (noted $\theta_{j}$).
Both distributions, the topic's specific language (bag of words) and the topics for each document, are not known beforehand. 

As mentioned before, the number of topics to be mined is a parameter of the model and varies greatly between different data sets and applications.
For example, one could estimate the amount of major topics and treat each topic model as the basis for a separate FAQ.
Another possibility is to decide on a fixed number of questions to include in the FAQ, to mine an according number of ``question models'' and to select the most representative Q\&As for each.
Since our evaluation shows that our approach is able to filter out unfocused models, the exact number of ``actual'' topics is not required.
Also, as it usually only requires a few minutes to generate FAQs, several strategies can be tried.

The objective of LDA is to approximate both distributions by maximizing the overall probability of the model, i.e., by maximizing the following function.
$K$ is the number of topics, $M$ is the number of documents, $N_j$ is the number of word instances (also called ``tokens") for document $j$, $W_{j,t}$ identifies the word for each instance $t$ in a document $j$, and $Z_{j,i}$ to which topic this word instance is assigned:
\begin{equation}
\label{eq:lda}
P(\theta, \varphi) =
\prod_{i=1}^K
P(\varphi_i) \prod_{j=1}^M P(\theta_j) \prod_{t=1}^{N_j}
P(Z_{j,t}|\theta_j)P(W_{j,t}|\varphi_{Z_{j,t}})
\end{equation}
\vspace{0cm}

Latent Dirichlet allocation also uses Dirichlet priors $\alpha$ and $\beta$ on $\theta$ and $\varphi$ respectively, which is out of the scope of this presentation ($P(\varphi_i)$ and $P(\theta_j)$ are assumed to be 1).
The approximation of the model parameters ($\theta, \varphi$) is generally done using a technique called Gibbs sampling \cite{Sammut2011}.

\medskip
For our implementation of the approach, we use \textit{Mallet} ("MAchine Learning for LanguagE Toolkit"), a Java library developed by McCallum and colleagues at the University of Massachusetts Amherst \cite{McCallumMALLET}.
It contains several text mining tools, including an implementation of LDA.
An important feature of the Mallet implementation of LDA is that it automatically adjusts model parameters (Dirichlet priors $\alpha$ and $\beta$) during optimization to further improve the accuracy of the mined topics.

\subsection{Resulting Data}
\label{mining-results}

As described, LDA extracts a set of topics, each of which is a candidate to be transformed into a FAQ.
Each LDA topic model represents a common ``conversation topic'' that is frequently discussed in the mailing list.
Each document is given a probability distribution over topics, i.e., for every document, LDA outputs a relative likelihood percentage of relating to a mined topic.
For example, one document might have a likelihood of 0.85 to relate to \textit{topic 2} and of 0.15 to relate to \textit{topic 7} (the sum is always 1).
In this case, the document mostly refers to \textit{topic 2} and is a good candidate to be included in the FAQ representing this topic.

\emph{Filter (Removing associations with low probabilities):}
We associate a document $j$ to a topic $i$ only if the likelihood $\theta_{j,i}$ is higher than a certain threshold.
We set this threshold to 0.25, meaning that we consider only relationships between documents and topics when at least one fourth of the document content relates to the topic.
If a document is not associated to at least one topic ($\theta_{j,i}<0.25$ for each i), it is definitely discarded, i.e., it appears in no FAQ at all.

%% file: faq-assembly.tex
The process of turning mined topic models to corresponding FAQs consists of three phases.
The first phase constitutes a FAQ by selecting Q\&A pairs from the conversations associated with a topic.
The second phase discards certain mined topics that seem unfocused. 
Finally, the questions of the remaining FAQs are ordered to maximize readability.

\subsection{Question and Answer Selection}
\label{qa-selection}

For selecting questions and answers to include in the FAQ, we compute a similarity metric between each entry in a conversation and the associated topic model.
The metric co-relates the weighted bag of words ${\bf t}$ given by LDA (see Section \ref{LDA}) with a word vector ${\bf e}$ representing the entry's normalized term frequencies. 
We use cosine similarity for comparing the two vectors, which is defined as follows:
\begin{equation}
\cos ({\bf t},{\bf e})= {{\bf t} {\bf e} \over \|{\bf t}\| \|{\bf e}\|} = \frac{ \sum_{i=1}^{n}{{\bf t}_i{\bf e}_i} }{ \sqrt{\sum_{i=1}^{n}{({\bf t}_i)^2}} \sqrt{\sum_{i=1}^{n}{({\bf e}_i)^2}} }
\end{equation}

A high cosine value indicates that an entry is closely related to the topic model and thus a good candidate for being a question or answer in the final FAQ on this topic.
Hence, we will use it in both question and answer selection.

\medskip
An entry in a conversation is selected as a \emph{good question} if it satisfies the following three conditions:
(a) it is the first one in the conversation;
(b) its cosine value with respect to the topic model (the FAQ theme) is relatively high (we found 0.15 to be a good threshold, see Section \ref{parameters}); and
(c) it is short enough. 
The rationales for the latter point are as follows.
On the one hand, for 15 popular software development FAQs we analyzed (1,000 questions in total), the average question length is 75 characters, which is rather short.
On the other hand, we assume that long questions are either too specific (complex scenarios often need many words to be explained) or are imprecisely formulated.
We use a threshold of 300 characters which encompasses most questions from existing FAQs and gives very good results in practice for automated FAQ extraction (see Section \ref{qualitative-evaluation}).
Selected questions may still not be part of the final FAQ if they are not clearly answered, i.e., if we cannot select an entry that meets the answer selection criteria.

In the standard FAQ format there is just one answer per question.
In Internet conversations, there is an arbitrary number of replies of varying relevance and quality to each question.
So, the best reply to each question has to be determined.
The key to estimating an answer's relevance is the language that is used.
An expert usually employs a significant amount of domain-related terminology; for example, he might point to several software library features or describe a usage scenario in details.
In comparison, other participating users might not have such understanding of the domain and therefore use a less specific language.
As the domain's terminology is modeled in the topic's bag of words, to select the best answer, we use again the cosine similarity between a reply and the related topic's bag of words.

The best reply is expected to be the one with the highest correlation in terminology with respect to the topic model, which we measure through cosine similarity.
If the highest correlation is still low, we discard the selected question, since it misses a precise and thorough answer (if the cosine similarity is lower than 0.15, see Section \ref{parameters} for an explanation on how we determine this).
 
\subsection{Topic Selection}
\label{topic-selection}

In this phase, FAQs constructed from the mined topics are filtered if too few question-answer pairs were selected in the previous phase.
The rationale for this is that sometimes topic models have no clear focus, which results in low-quality FAQs.
For instance, if a mailing list discusses three major topics (e.g. development, maintenance and customer support), but LDA was set up to mine 4 models, one of the models contains all conversations that are ``off-topic'' with respect to the 3 major topics.
Those conversations do not discuss a common topic (there is no ``actual'' forth topic), which induces a more vague and unfocused bag of words compared to the focused topics.
Consequently, many conversations won't yield question-answer pairs surpassing our selection filters, due to a lack of semantic correlation. Based on this characterization, we define the following filter.

\emph{Filter (Removing unfocused topics):}
To remove unfocused topics as a whole, we require a minimum threshold on the amount of selected questions ($Q\&A$) compared to the number of initial conversations in the topic:
\begin{equation}
\frac{|{Q\&A}|}{|{Conversations}|}>threshold
\end{equation}

\subsection{FAQ Ordering}

Finally, inside an FAQ, the Q\&As themselves are ordered by their relevance to the FAQ theme, as given by the harmonic mean between the question's and the answer's cosine similarity with the topic.
As opposed to the arithmetic mean, the harmonic mean requires both question and answer relevance to be high to yield a high value, i.e. it ensures that for the top Q\&A pairs, both questions and answers, are of high quality.

The resulting FAQs are meant to be validated by experts before being published.
This means that some Q\&As may be finally discarded, but more importantly, Q\&As may be slightly reformulated or detailed to improve the quality of the final FAQ.

%% file: qualitative-evaluation.tex
The goal of the proposed approach is to automatically create FAQs that are meant to be published as official software documentation.
Therefore, we have constructed an evaluation setup that answers to the question: ``Are our generated FAQs considered valuable by software project leaders?''.
For 32 popular open-source projects, we invited the top committers to go through a guided evaluation interface, where they were first asked to select relevant questions from an automatically created FAQ and then to evaluate the respective answers.
From the results, we estimate the relevance and correctness of our mined topics, automatically selected questions and automatically selected answers.

\subsection{Data basis and setup}
\label{data-basis}

To construct a representative set of software development mailing lists, we chose 50 of the 200 most popular projects found at \href{http://www.ohloh.net}{ohloh.net}, an open-source software directory.
Specifically, we chose those projects for which we could retrieve the corresponding mailing lists from the archives at \href{http://markmail.com}{markmail.com}.
From these, we excluded conversations published before 2009, a compromise between having only up-to-date questions and having enough data for making significant observations.
In total, our data contains 310,000+ messages of 70,000+ conversations from 50 lists.

For each selected project, we also obtained a list of the most active committers from Ohloh  (those who made most contributions to the source code repository), to whom we sent a link to a website-based review system for the project's generated FAQ.
In total, 500 invitations were sent out, nearly equally distributed among 32 projects (for some, less than 15 committers were contactable).
I.e., all reviewers were at least among the 25 most active contributors to their respective project.
18 of the 50 projects were not included in the expert review because the associated topic models were automatically removed (see Section \ref{topic-selection}) or, in fewer cases, no contact data could be found.

For generating the FAQs to be reviewed, we mined a single topic model per mailing list to get the most frequent topic and we applied our FAQ assembly techniques to this topic.
In other terms, this evaluation does not assess the importance of the number of topics, but rather the impact of processing and FAQ assembly techniques to select good Q\&As.

\subsection{Review process}

After a short explanation of our approach, the reviewer is presented the 40 highest scoring questions that our approach extracted from the associated mailing list 
(in terms of cosine similarity, see Section \ref{qa-selection}) and is asked to select all questions he considers to be worth publishing.
Since the expert is further asked to review one proposed answer for each selected question, we advised him a 15 questions maximum to prevent demotivation and breaking off the review process during the answer validation phase\footnote{For 15 questions, a review takes 10-15 minutes on about 20 pages.}.
At this point, the answers that our system has chosen are not given, in order to get the reviewer opinion on the questions only.
Additionally, the questions are not ordered by cosine similarity, but by their length.
First, shorter questions are more motivating to start with.
Second, this length based ordering avoids a cosine-based bias, which would inhibit an evaluation of this crucial score.

Once the reviewer has completed the question selection phase, each selected question is presented on a separate page along with the answer selected by our approach. 
There are three options available to the reviewer: 1) to validate the answer (i.e., accept the answer as is); 2) to modify the answer text (and question text, if needed) and 3) to discard the question, because the presented answer does not match 
his expectations and cannot be corrected effortlessly.

Eventually, the reviewer can leave a comment and download a XML or HTML version of the whole FAQ he has just rolled out.
On \href{http://faqcluster.com}{faqcluster.com}, we present the validated FAQs that experts agreed to publish on our own website.
For the evaluation, we only considered those reviews that are completed seriously, i.e., reviews with at least 5 questions selected and all corresponding answers evaluated.

\subsection{Survey results}

Table \ref{table-reviews} provides the most important results of this evaluation.
The first four rows give general information about the participation and how many projects got reviewed at least once.
The data shows that 24\% of all invited reviewers actually visited the review page.
Since we did not give much information about the subject in the emails, we conclude that they either not participate in surveys in general or the email was not read at all (many committers finished contributing up to 10 years ago so they may not further be interested in their projects or the email address may not be in use anymore).
When visiting the website however, 26 experts (23\%) took the time to complete the whole review process.

\input{table-qualitative}

The next section of the table is concerned with the quality of the selected questions and answers, how many of them were edited, and, if answers got discarded, for which reasons.
To be able to also compare selected questions with questions intentionally not selected, we split the 40 questions into 5 pages of 8 questions and only consider the questions of those pages visited by the reviewer (if he/she clicked on the link to this page).
For an average of 35 questions looked at, 16 were selected, which indicates that most of the reviewers were able to find a sufficient amount of questions (they surpassed the proposed limit of 15) and at least 46\% of all questions selected by our approach are relevant for a FAQ.
18\% of all questions then were discarded because the provided answer was insufficient - in most cases either wrong (45) or imprecise (22).
In other terms, 82\% of answers were of sufficient quality for a FAQ (at least after slight modifications).
Finally, comparing selected questions to intentionally skipped ones also allows us to measure the correlation between cosine similarity and questions selection.
The average selected question has a cosine value of 0.3359, all others only 0.2756, i.e., the higher the cosine similarity, the significantly lower the probability of irrelevant questions (99.9\% confidence).

To sum up, most reviewers were able to select more than a dozen questions, some even significantly more, and almost half of all questions selected by our approach were considered relevant.
The evaluation also shows that it is often challenging to find an appropriate answer: only 77\% of answers were correct or already sufficiently precise.
This shows that having a statistical ``surface'' understanding of text is not precise enough in all cases.
This may also indicate that there is sometimes no complete solution given in one response, e.g., when the answer is distributed across several replies of the same discussion.

The last section of the table gives information about message length and editing amount.
For instance, the selected questions have an average length of 174 characters, and the selected answers 442 characters.
This confirms that expert tend to prefer short questions and that answers are generally longer, as observed in real-life FAQs.
When reformulated, the selected questions have a mean Levenshtein distance of 58 characters, i.e., not much effort is required in making them publishable.
The answers validated (marked as correct by experts), yet reformulated, were changed by an average amount of 44 characters, i.e. 10\% of changes.
Interestingly, the experts reformulated the questions much more than the answers.
Unfortunately, there is no way to know whether this is because of the text length (it is easier to reformulate short texts), because experts care more about question clarity than answer clarity or because answers in mailing lists are generally of a higher standard than question and thus do not need much revision.

The personal comments received in the feedback text furthermore indicate that some open source leaders have a clear interest in automatically generating FAQs. 
For instance, a reviewer said that \emph{``Your project seems quite interesting, and automatically generating a meaningful repository of information out of the mess that 
mailing lists are seems to me like a very interesting development in NL processing technology.''}.
To sum up, this evaluation shows that our approach is able to generate software development FAQs, 
which only require a short review before they can actually be published.

%% file: table-qualitative.tex
\begin{table}[!t]
\renewcommand{\arraystretch}{1.3}
\caption{Qualitative Evaluation Results.}
\label{table-reviews}
\centering
\newcolumntype{C}[1]{>{\centering\arraybackslash}p{#1}}
\begin{tabular}{p{4.2cm}||C{1.05cm}||C{2.1cm}}
\hline
\bfseries Category & \bf Amount & \bf Relative \\ \hline
\hline
Selected open-source projects  & 32 & - \\
Review invitations sent  & 500 & - \\
Website visited once & 118/500 & 24\% \\
Reviews completed & 26/118 & 23\% \\
Reviewed projects & 17/32 & 53\% \\
\hline
\hline
Q\&A to be reviewed & 1280 & 40 / rev.  \\
Questions viewed  & 905/1280 & 71\%, 34.8 / rev. \\
Questions selected  & 413/905  & 46\%, 15.9 / rev. \\
Questions modified & 29/413 & 7\% \\
Mean cosine selected questions & 0.3359 & -  \\
Mean cosine not selected questions & 0.2756 & -  \\
\hline
Correct Answers & 337/413 & 82\% \\
~~~~Answers as-is & 259/337 & 77\% \\
~~~~After modification & 78/337 & 23\% \\
Answers discarded & 76/413 & 18\% \\
~~~~Wrong answer & 45/76 & 59\% \\
~~~~Imprecise answer & 22/76 & 29\% \\
~~~~Other & 9/76 & 12\% \\
\hline
\hline
Selected questions mean length ($l_q$) & 174 & -  \\
Selected answers mean length ($l_a$) & 442 & -   \\
Mean Levenshtein dist. quest. / $l_q$  & 58/174 & 33\% \\
Mean Levenshtein dist. answ. / $l_a$ & 44/442 & 10\% \\
\hline
\end{tabular}
\vspace{-.5cm}
\end{table}

%% file: quantitative-evaluation.tex
When selecting Q\&A pairs in Section \ref{faq-assembly}, we rely on topic models (specified as bag of words) to distinguish between good questions and answers and those which are irrelevant for FAQs.
This second evaluation estimates the capabilities of LDA to provide such models and the performance and importance of our pre- and post-progressing steps in improving them with respect to our requirements.
In particular, we provide evidence showing that:
(a) LDA can  actually make sense of software-related discussions using appropriate pre-processing (Section \ref{preimp}); 
(b) the post-processing of the LDA's results presented significantly improves the quality of the generated FAQs (Section \ref{fimp}); and 
(c) the approach's parameter settings are set and optimized in a systematic manner (Section \ref{parameters}).

\subsection{Evaluation Setup}

The evaluation is set up to investigate LDA's performance in constructing topic models from software development discussions and to see how those models can be improved by our filters.

To do so, we first merge all 50 mailing lists of our evaluation data (see Section \ref{data-basis}) into one single data set, 
leaving no explicit distinction between conversations from different sources.
Second, we configure LDA to create exactly 50 topic models from the merged conversation set and expect 
to have each model correlating with exactly one mailing list.
Finally, we apply several metrics to measure the success in reconstructing the original mailing lists.

The metrics we use for measuring success are precision and recall.
For each mined topic, we identify the mailing list from which most of the topic's associated conversations originate and call it the \emph{main mailing list}.
Precision measures the percentage of the topic's conversations that belong to that main list, i.e., a precision of 1 means that no conversations from other lists are included.
Recall measures the percentage of the list's conversations being included in the model, i.e., a recall of 1 means that the entire mailing list is included in the topic model.
Monitoring both values is important since, for example, a precision of 1 could be achieved by having only 1 conversation; 
obviously, this model would be useless.

We use the approximately same amount of conversations for each subject mailing list to avoid the effects of an unequal distribution of mailing list sizes on the results.
For instance, with a mailing list being too large in proportion, LDA could create multiple topics from this list, while several smaller lists would be merged to one model (the amount of topics is fixed beforehand).
The smallest mailing list contains 214 conversations from 2009 to 2011, the largest 10,966; the median is 573,
hence we limit the amount of conversations used in the experiment to 600 for each mailing list.

\subsection{Evaluation of FAQ Assembly}
\label{fimp}

Table \ref{table:concepts} shows precision and recall for certain mined topics.
For each mined topic, precision and recall are computed with respect to the closest corresponding mailing list (given in brackets).
Values are given for ``PP'' models with pre-processing only and ``FAQs'' with both pre- and post-processing (FAQ assembly) switched on.
The rows present best and worst topic models with respect to the highest harmonic mean between precision and recall after FAQ assembly (also known as the F-score).

For instance, topic \#8 initially contains all conversations of the NHibernate mailing list (a recall of 1). For this topic, the FAQ question and answer filtering increase the precision to 1 (meaning that only conversations from NHibernate remain after filtering). On the contrary, topic \#25 corresponds to no mailing lists at all (a recall of 0.0611 after LDA), but is filtered out by our topic filter (indicated by an ``x'' in the FAQ column).

When comparing the bags of words in Figure \ref{fig:bagofwords}, it is obvious that topic \#31 models an email terminology perfectly corresponding to the email client Mutt.
Topic \# 36, however, does not seem to relate to a specific software package at all, but to software development in general; it is rather accidental that Firebug is the main mailing list 
(only 7.69\% of all conversation originate from the Firebug list).

This table shows two interesting points. First, all unfocused topics are filtered out by the post-progressing techniques (14/50 are filtered out, including 9 of the 10 worst topic models).
Second, for the focused topics, the precision is always increased after FAQ assembly; the average precision jumps from 0.5711  to 0.7634.

\emph{To sum up, if a generated topic model represents one meaningful topic, one can be confident that our FAQ assembly techniques improve the precision, i.e. the focus of extracted Q\&As, or otherwise filter them out.}

\input{table-topics}
\input{figure-words}
\input{table-topics-summary}

\subsection{The Pre-processing Heuristics Improve LDA's Models}
\label{preimp}

Table \ref{table:concepts-summary} shows average and median precision, recall, and total number of Q\&As for all mined topics.
The difference with Table \ref{table:concepts} is that it also shows the precision and recall of using basic LDA on raw data.
For instance, as shown in the row labeled "Median", LDA on raw data has a median recall of 0.5534 and a median precision of 0.4675.

LDA on top of pre-processed data (column "PP") has a precision of 0.7717 and a recall of 0.6275, respectively.
Compared to raw LDA, this is a significant increase in both precision and recall.
In the conversation column, one can see that pre-processing yields more conversations related to topic models (517 instead of 447), which is due to having more focused topics models resulting in the topic containment probabilities $\theta_{j}$ of a conversation's major topics (see Section \ref{LDA}) being generally higher than the threshold defined in Section \ref{mining-results}.

\emph{This shows that our techniques to filter noise specific to software development, such as stack traces and debug logs, do increase LDA's capability to grasp the actual semantics of software mailing lists.}

The results of FAQ columns are identical to Table \ref{table:concepts}, but in this table they contribute to showing our overall goal: to improve the precision of the approach. In general, we prefer having fewer Q\&As of high quality rather than many Q\&As of medium preciseness. The column ``Precision'' of Table \ref{table:concepts-summary} exactly shows this, we are able to constantly increase
 the precision first with the pre-processing, then with the FAQ assembly, while still retaining a reasonable number of Q\&As at the end (118 Q\&As per mined topic in average, to be compared with the 600 input conversations).

\subsection{Importance of the Approach Parameters}
\label{parameters}

\input{table-parameters}

Finally, since our approach consists of many heuristics, filters and thresholds, we are interested in understanding to what extent they contribute to the performance of the system.
We use the same setup as in the previous sections, but instead of using only the optimized configuration, we run the whole approach with a variety of parameter settings.
For each parameter, we select the optimized configuration we obtained from an exploration of the parameter value space (with respect to qualitative and quantitative evaluation, e.g., not too much filtering) and then change the parameter value to observe the effects on the evaluation metrics.

In Table \ref{table-parameters}, each line represents one of the approach's parameters being changed from the optimized settings.
The given evaluation metrics refer to the performance of the second evaluation task, i.e., the reconstruction of mailing lists including the FAQ assembly phase.
As already seen, turning off the pre-processing yields a significant decrease of performance; the results indicate that each heuristic contributes to a higher precision and recall.
However, precision values are still better than for the LDA-only models from Table \ref{table:concepts-summary} (column ``PP'') since our FAQ assembly techniques are able to dampen the negative effects.
Note that disabling the LDA model threshold also leads to a decrease in precision.
Increasing it accordingly leads to a higher precision (0.95), however, it has a strong impact on the number of generated FAQs (then only 15 out of 50 FAQs are kept instead of 36).
Changing the cosine threshold on question and answer selection shows the correlation with our metrics; the higher the threshold, the higher the precision and the lower the recall.
The same is evident for the FAQ selection, where a higher threshold also means more precision since unfocused FAQs are filtered out.

\subsection{Limitations of the Quantitative Evaluation}

In this section, we have shown that our approach is able to recognize topics that are hidden in data. We know those topics because we seed them artificially by merging different mailing lists. While this works on our particular dataset, it may be the case that those mailing-lists are actually very much different, with a clearly orthogonal terminology to describe them. We hope that replication and extension of our work will introduce new datasets to improve the confidence of the topic mining results.

For the sake of creating FAQs, there may be some topics that crosscut different mailing lists. With the same evaluation setup, one could use tagged data instead of completely disjoint lists in order to evaluate the emergence of cross-cutting topics using our approach.

Finally, it is clear to us that this automated evaluation does not assess whether the mined questions and their corresponding answers are good with respect to relevance to the software documentation, the clarity, the style, etc. Those quality attributes were evaluated by the user study presented in Section \ref{qualitative-evaluation}.
However, in this work, we assume that a good precision in reconstructing merged mailing lists is indirectly linked with good Q\&As.
Based on this assumption, we have calibrated the various thresholds and parameters of the approach in a fully automated manner before setting up the user study.
The fact that the ratio of correct questions and answers (as assessed by experts) is high gives support to our core assumption.

%% file: table-topics.tex

\begin{table}
\renewcommand{\arraystretch}{1.2}
\caption{Reconstructing 50 Mailing-Lists with LDA.
The FAQ assembly techniques improve the precision and filter out unfocused topic.}
\label{table:concepts}
\centering
\newcolumntype{C}{>{\centering\arraybackslash}X}%
\begin{tabularx}{\columnwidth}{p{2.37cm}||C|C|C|C}
\hline
\bfseries Topic Model (Main List) & \multicolumn{2}{c|}{\bf Recall} & \multicolumn{2}{c|}{\bf Precision}  \\ 
\hline
\bf  &  \bf PP & \bf FAQ &  \bf PP & \bf FAQ  \\
\hline \hline
\bf Best topics  &&&&\\
\#8 (NHibernate) &  1.0 & 0.3743 & 0.9632 & 1,0  \\
\#46 (D-Bus) &  0.8564 & 0.3479 & 0.8167 & 0.9408  \\
\#31 (Mutt) &  0.86 & 0.32 &  0.8487 & 0.96  \\
\#44 (Hudson) &  0.6767 & 0.29 &  0.7778 & 0.9305  \\
\#17 (Log4J) &  0.8063 & 0.3099 &  0.4761 & 0.8627 \\

\bf Worst topics  &&&&\\
{\#20 (Firebug)} &  0.075 & x &  0.1546 & x  \\
{\#15 (Hudson)} &  0.03 & x  &  0.0957 & x   \\
{\#49 (Mediawiki)} &  0.0183 & x &  0.1134 & x \\
{\#36 (Firebug)} &  0.015 & x &  0.0769 & x   \\
{\#25 (Greasemonkey)} &  0.0611 & x &  0.14 & x   \\

\hline
\bf Std Deviation  & 0.1712 & 0.0763 &    0.2817 &  0.1911\\
\bf Average  & 0.6071 & 0.1829  & \bf 0.5711 & \bf 0.7634   \\
\bf Median & 0.7717 & 0.1858   & \bf 0.6275 & \bf 0.8261  \\

\end{tabularx}
\end{table}

%% file: figure-words.tex
\begin{figure}[!t]
\renewcommand{\arraystretch}{1.4}
\begin{scriptsize}
\begin{tabular}{l}
\textit{Top 20 terms for high-score topic \#31 (Mutt)} \\
mail (0.079), mutt (0.078), message (0.063), messages (0.035), list (0.031), \\ email (0.030), folder (0.028), set (0.025), gmail (0.020), hook (0.019), send \\ (0.019),  imap (0.016), attachment (0.015), subject (0.014), mime (0.013) \\
\\
\end{tabular}
\begin{tabular}{l}
\textit{Top 20 terms for low-score topic \#36}\\
code (0.124), write (0.032), dont (0.029), make (0.025), feature (0.024), \\ simple (0.023), idea (0.021), add (0.020), api (0.020), solution (0.020), \\ implement (0.017), question (0.016), part (0.015), work (0.014), documentation (0.014)
\end{tabular}
\end{scriptsize}

\caption{Bags of words representing a focused topic \#31 (Mutt) and an unfocused one \#36. The first one contains a very specific terminology, the second one only contains generic software development terms.}
\label{fig:bagofwords}
\end{figure}

%% file: table-topics-summary.tex

\begin{table*}
\renewcommand{\arraystretch}{1.2}
\caption{Conformance between mined topics and corresponding mailing lists. ``Raw'' columns give results for LDA models without pre-processing, ``PP'' columns give LDA results after pre-processing, FAQ columns give results after all filtering phases. 
Pre-processing improves the performance of LDA in both precision and recall.}
\label{table:concepts-summary}
\centering
\newcolumntype{C}{>{\centering\arraybackslash}X}%
\begin{tabularx}{\textwidth}{p{4.1cm}||C|C|C|C|C|C|C|C|C}
\hline
\bfseries Topic Model (Main Mailing List) & \multicolumn{3}{c|}{\bf Recall} & \multicolumn{3}{c|}{\bf Precision} & \multicolumn{3}{c}{\bf \# Q\&As} \\ 
\hline
\bf  & \bf Raw & \bf PP & \bf FAQ & \bf Raw & \bf PP & \bf FAQ & \bf Raw & \bf PP & \bf FAQ \\
\hline \hline
\bf Standard Deviation & 0.1504 & 0.1712 & 0.0763 &  0.3235 &  0.2817 &  0.1911 & 233 & 249 & 39 \\
\bf Average & \bf  0.4715 & \bf 0.6071 & 0.1829 & \bf 0.4996 & \bf 0.5711 &  0.7634 & 471 & 499 & 118  \\
\bf Median & \bf  0.5534 & \bf  0.7717 & 0.1858  & \bf 0.4675 & \bf 0.6275 &  0.8261 &  447 & 517 & 111 \\
\hline
\end{tabularx}
\end{table*}

%% file: table-parameters.tex
\begin{table*}
\renewcommand{\arraystretch}{1.3}
\caption{The impact of the approach parameters on the metric medians. The first line is the baseline corresponding to the optimized configuration. The other lines read as a comparison against it.}
\label{table-parameters} 
\centering
\newcolumntype{P}[1]{>{\centering\arraybackslash}p{#1}}
\newcolumntype{C}{>{\centering\arraybackslash}X}%
\begin{tabularx}{\textwidth}{p{5.0cm}|C|C||C|C|C}
\hline
\bfseries Parameter differing from default setting & \bf Default Value & \bfseries New Value & \bf FAQ Recall & \bf FAQ Precision & \bf \# Q\&As \\ \hline
\hline
Optimized Setting & -  & -  & 0.1858 &  0.8261 & 111\\
\hline
Heuristic 1 (\ref{rep-display}) & on & off & 0.1683 & 0.7595 & 123\\
Heuristic 2 (\ref{rep-display}) & on & off & 0.1683 &  0.7333 & 111\\
Heuristic 3 (\ref{internal-representation}) & on & off &  0.1467 & 0.6949 & 129\\
Heuristic 4 (\ref{internal-representation}) & 800 & off &   0.1564 &  0.721 & 117\\
Heuristic 5 (\ref{internal-representation}) & on & off &  0.1823 &  0.7893 & 114\\
All Heuristics (\ref{rep-display} \& \ref{internal-representation}) & on & off & 0.1311 & 0.6269 & 134\\
\hline
LDA Model Threshold (\ref{mining-results}) & 0.2 & 0.05 & 0.26 & 0.7484& 224\\
LDA Model Threshold (\ref{mining-results}) & 0.2 & 0.4 & 0.1421 & 0.9512 & 78\\
\hline
Q/A Cosine Threshold (\ref{qa-selection}) & 0.15 & off & 0.5534 & 0.4675 & 447\\
Q/A Cosine Threshold (\ref{qa-selection}) & 0.15 & 0.1 & 0.2308 & 0.7783 & 160\\
Q/A Cosine Threshold (\ref{qa-selection}) & 0.15 & 0.2 & 0.1392 & 0.8939 & 87\\
FAQ Selection Threshold (\ref{topic-selection}) & 0.1 & off & 0.1612 & 0.6942 & 116\\
FAQ Selection Threshold (\ref{topic-selection}) & 0.1 & 0.2 & 0.201 & 0.9135 & 139\\
FAQ Selection Threshold (\ref{topic-selection}) & 0.1 & 0.3 & 0.3455 & 0.9342 & 152\\
\hline
\end{tabularx}
\end{table*}

%% file: related-work.tex
Question answering systems have a long tradition in artificial intelligence research.
Certain systems are designed to find the most appropriate answer in a database of question/answer pairs (e.g. \cite{Burke:1997:QAF:896178,jijkoun2005retrieving}).
Others use the world wide web as reservoir of answers (e.g. \cite{kwok2001scaling,Harabagiu2006,Celikyilmaz2009}).
Certain authors only focus on finding similar questions to those asked by users (e.g. \cite{jijkoun2005retrieving,jeon2005finding}).
Compared to this previous research, we do not work with a structured set of Q\&As. In our input data, we know neither the questions nor the corresponding answers.

Jeon et al. \cite{jeon2006framework}, Surdeanu et al. \cite{surdeanu2008learning} and others \cite{Shrestha2004,cong2008finding,hong2009classification} use a variety of approaches for learning to detect questions or correct answers from mailing lists and forums.
Their approaches differ from ours as following: first they do not try to group Q\&As by topics as we do, second they do not address the details specific to software discussions. The ``lexical chasm'' refers to the fact that an answer contains a terminology that is missing the question. Interestingly, when we measure the similarity of the question and the answer to the topic's bag of words directly, we assume that the topic contains the terminology of both the question and the answer.
The selected Q\&As with high cosine similarity are very close to it.
In other terms, compared to this related work that finds ``questions'' and ``correct answers'', to a certain extent, we find Q\&As that are also ``clear'' and ``precise'' (if the topic's bag of word is focused and relevant).

Gottipati et al. recently presented \cite{Gottipati2011} an approach to finding answers of software development questions in a manner that is similar to search engines.
First, their system performs a classification of software forum posts into 7 categories: questions, answers, question/answer clarification, question/answer feedback and junk.
Second, it offers a standard search engine query interface which uses the categorization to improve the search performance and to only display answers as results.
Our approach is similar with respect to identifying relevant answers.
While they construct a search engine, we try to infer relevant and precise topics and questions from data, which is not in their scope at all.

Celikyilmaz et al. \cite{Celikyilmaz2010} and  Liu et al. \cite{Liu2010} also use LDA in the context of question answering systems.  
Celikyilmaz et al. \cite{Celikyilmaz2010} use LDA on possible answers to a given question. On the contrary, we use LDA on the whole mailing-list to identify the frequent topics and the bag of words representing them. 
While it makes sense to build an LDA model on the web, it does not make sense to build a LDA model on a small dataset such as the replies to an initial question (usually less than a dozen replies). 
Liu et al. \cite{Liu2010} use LDA to predict the best potential answerers to a new question;  their LDA models capture different topics an expert is interested in. 
In the domain of software engineering, Hindle et al. \cite{Hindle2011} cluster commit-log comments and others cluster source code \cite{Grant2010,Thomas2011,Kelly2011}, all with LDA.
On the contrary, our LDA model captures topics of mailing lists.

The concept of applying data-mining techniques to FAQs has been explored by Ng'ambi \cite{NgAmbi:2002:PUQ:581506.581521}.
His system guesses the next question that the user would ask (``pre-empting'' user questions).
We use data mining in a different context.

Different authors have explored different ways of automatically inferring software documentation.
The approaches vary in terms of both input information used to generate documentation and the kind of generated documentation artifacts.
Buse et al. \cite{Buse:2008:ADI:1390630.1390664}  analyze method bodies to infer API documentation related to exceptions.
Long et al. \cite{Long:2009:AHV:1595696.1595727}  also use source code to infer relevant links between API elements to improve developer's awareness.
Bruch et al. \cite{Bruch2010} extract subclassing directives also from source code to assist developers in correctly extending object-oriented frameworks.
Sridhara \cite{Sridhara:2010:TAG:1858996.1859006} uses method signatures to infer meaningful natural language sentences to be included in a method's summary.
In contrast, our approach infers FAQs, which is another frequently used kind of software documentation.
Furthermore it uses mailing lists and forums as input and not source code.

Mailing lists have been used to infer social networks \cite{bird2006mining} and their relations with software development. 
The approach uses structured email headers (e.g. the ``From'' and ``To'' headers). On the contrary, we use the content of mailing lists and forums; the latter contain a large amount of noise 
not present in email headers.

%% file: conclusion.tex
In this paper, we presented an approach to semi-automatically extract FAQs from software development forums and mailing lists.
The three main components of the approach are:
(a) applying several pre-processing heuristics to remove noise from the raw data and to prepare it for further steps and the final display;
(b) using latent Dirichlet allocation (LDA) to automatically extract an arbitrary amount of topic models from the pre-processed data, each of which may serve as the basis for a topic-specific FAQ;
(c) using the model information for selecting and ordering most relevant questions, selecting the most relevant answer and estimating the FAQ quality.

We conducted both a qualitative and quantitative evaluations. We asked top contributors to 50 major open source projects to review generated FAQs corresponding to their project.
The results confirm that the quality of the generated FAQs is promising:
almost half of all questions we proposed were considered relevant and 82\% of the automatically selected answers were correct.
For the quantitative evaluation, we extracted FAQs from a set of merged mailing lists to prove our ability to identify the most important topics.
We observed a strong correlation between the hidden structure (the mailing lists) and the mined topics (the FAQs).
Furthermore, we were able to show that our filtering techniques specific to software development and FAQ generation could significantly improve the performance of topic mining compared to basic LDA.
The replication data can be found on \url{http://faqcluster.com/replication}.

The work presented in this paper lies in a wider area of research dedicated to improving the accessibility of software development data and knowledge.
Since we are the first to generate this kind of documentation, we are convinced that there is much room for creativity and optimization in order to improve the quality and the completeness of the generated FAQs.
From a wider perspective, an important future work consists of checking, if not inferring, the semantic links between different forms of documentation: API documentation, FAQs, tutorials, etc.